\documentclass[global,twocolumn]{svjour}
\usepackage{graphicx}
\journalname{{\protect Applied Physics B manuscript No. 9004 B \hspace*{20cm}}}
\makeatletter
\def\lambdabar{\protect\@lambdabar}
\def\@lambdabar{%
\relax
\bgroup
\def\@tempa{\hbox{\raise.73\ht0
\hbox to0pt{\kern.25\wd0\vrule width.5\wd0
height.1pt depth.1pt\hss}\box0}}%
\mathchoice{\setbox0\hbox{$\displaystyle\lambda$}\@tempa}%
{\setbox0\hbox{$\textstyle\lambda$}\@tempa}%
{\setbox0\hbox{$\scriptstyle\lambda$}\@tempa}%
{\setbox0\hbox{$\scriptscriptstyle\lambda$}\@tempa}%
\egroup
}
\makeatother

\begin{document}

\title{Loss and heating of particles in small and noisy traps}

\author{Carsten Henkel \and Sierk P\"otting \and 
Martin Wilkens} 

\institute{Institut f\"ur Physik, 
Am Neuen Palais~10, Universit\"at Potsdam,
D--14469 Potsdam, Germany,  
\\
\email{Carsten.Henkel@quantum.physik.uni-potsdam.de} 
}

\date{25 june 1999 / revised 29 september 1999}

\maketitle

\abstract{
We derive the life time and loss rate for a trapped  
atom that is coupled to fluctuating fields in the vicinity  
of a room-temperature metallic and/or dielectric surface. 
Our results indicate a clear predominance of near field effects
over ordinary blackbody radiation. We develop a theoretical
framework for both charged ions and neutral atoms with and
without spin. Loss processes that are due to a transition
to an untrapped internal state are included.
}

\titlerunning{Loss and heating in small and noisy traps}
\authorrunning{C. Henkel, S. P\"otting, and M. Wilkens}



\section{Introduction}

Particle traps enjoy great popularity for the preparation and  
manipulation of coherent matter waves. Prominent applications are  
the preparation of non-classical states of motion of a single ion
\cite{Wineland96},  
the realization of quantum gates in quasi-one dimensional ion traps
\cite{Wineland98c},  
the transfer of atoms through atomic wave guides
\cite{Ohtsu96b,Hinds98,Haensch98,Ertmer98}, and the  
preparation of quantum-degenerate gases in electromagnetic-solid  
state hybrid surface traps \cite{Ovchinnikov97b,Mlynek98b}. 
In all these applications, in order to  
truly benefit from the quantum mechanical effects, coherence of the  
matter waves and/or their internal degrees of freedom must be  
maintained as long as possible. Yet, with the physical components  
which provide the trapping potential being held at room  
temperatures, the maintenance of coherence seems highly non trivial  
as the temperature gradient between components and trap center
may well exceed $10^6{\rm K}/{\rm m}$. 
A careful study of the particles' coupling to the trap  
physical components, and the ensuing heating of the particles is  
therefore highly desirable.

In the past, the heating of single particles in small traps has  
been studied by a number of authors
\cite{Wineland75,Lamoreaux97,James98,Milburn98,Knight98,Wineland98}.
As these studies were mostly performed in the wake of the recent  
achievements in ion trapping and cooling, the focus in these  
investigations was on charged particles and their coupling to the  
surrounding metallic surfaces. In fact, before the advent of laser  
cooling, this coupling provided the dominant cooling mechanism for  
an ion cloud, say, as the low-frequency radiation of the ions  
couples quite efficiently to the lossy currents in the metallic trap  
components \cite{Wineland75}. 
Yet with the advent of laser cooling, temperatures of a  
few micro-Kelvin can be reached which are clearly below the  
components' temperatures, i.e.\ the particle-component coupling now  
leads to heating, and the trap ground state acquires a finite life  
time. Similar considerations may also be put forward for ultracold
neutral atoms
trapped in miniaturized traps though the couplings are different:
for paramagnetic atoms, e.g., they involve fluctuating magnetic 
rather than electric fields close to the trap components.

In this paper we derive the life time and loss rate for a trapped  
particle that is coupled to fluctuating fields in the vicinity  
of a room-temperature metallic and/or dielectric surface. The theory  
will be developed for both charged and neutral particles with and  
without spin, and loss processes that are due to a transition to an  
untrapped internal state will be included. A detailed derivation of  
previously published results \cite{Henkel99b} will also be given. 

An essential ingredient of the theory are cross-correlation functions for 
thermal electric and magnetic fields in a finite geometry.
These functions may be simplified for our purposes because 
the relevant field fluctuation frequencies are much lower
than the inverse time for light propagation 
from the trapped particle to the surface and back.
It is hence justified to calculate the fields
in the quasi-static limit, neglecting retardation effects. Differently
stated, the particle is subject to near field radiation leaking out
of the macroscopic trap components. An important consequence is
that the near field fluctuations are much stronger than those of the
well-known blackbody radiation. This implies larger than expected
heating rates, as recently pointed out by Pendry \cite{Pendry99b}.

The paper is organized as follows: in Sec.~\ref{s:model}, the model 
is presented in terms of a master equation. We identify the relevant
heating and loss rates. Sec.~\ref{s:ion} is devoted to trapped ion heating.
We give the electric field fluctuations above a flat metallic surface. 
In Sec.~\ref{s:spin}, heating and loss of a neutral particle 
with a magnetic moment is studied. 
The final Sec.~\ref{s:conclu} gives a summary and outlook. 
The appendixes contain technical material that is used in
the main text.


\section{The model: master equation and transition rates}
\label{s:model}
\label{s:master}

\begin{figure}[tbh]
\rotatebox{270}{%
\resizebox{!}{\columnwidth}{%
\includegraphics*{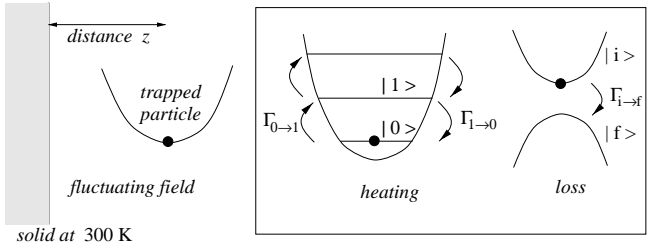}
}}
\caption[]{\label{fig:model}%
Left: trap in front of a flat surface. 
Right: heating and loss processes.}
\end{figure}

We present here our model for the particle trap and its 
environment (see fig.\ref{fig:model}, left part).
The model is sufficiently simple 
to allow for analytical calculations of the relevant heating and loss rates,
but also reflects a typical experimental geometry. 
We consider a single particle bound in a harmonic trap potential 
whose center is located at a distance $z$ from an infinite flat surface.
We consider that this distance is much larger than the
size of the particle's center-of-mass wave function.
In this regime, the overlap with the surface is negligible, and
the coupling to the surface is mediated via electromagnetic fields.
We also focus for simplicity on a single degree of freedom 
in the harmonic well. 

The heating of the particle is 
described by the transition rate $\Gamma_{0\to1}$
from the trap ground state $|0\rangle$ to the first excited state $|1\rangle$
(see fig.\ref{fig:model}, central part).
In subsection~\ref{s:heating}, 
such a `heating rate' is determined from a master
equation for the particle's motion in terms of harmonic-oscillator 
matrix elements, on the one hand, and the spectral density of a
fluctuating force field, on the other.

As a second application, we investigate loss processes in magnetic
or optical traps where only a subset of internal states is trapped
(see fig.\ref{fig:model}, right part). This model describes
magnetic traps, for example, where only low-field-seeking Zeeman
sublevels can be trapped. A loss process occurs when a fluctuating field 
induces a flip $|i\rangle \to |f\rangle$
of the particle's internal state. We assume that the particle
is then rapidly expelled and lost from the trap. The relevant loss rate
$\Gamma_{i\to f}$
is given in subsection~\ref{s:loss} in terms of internal matrix elements
for the particle's magnetic moment, on the one hand, and the magnetic
field fluctuation spectrum, on the other.

\subsection{Heating}
\label{s:heating}

As mentioned before, we focus on the heating of a single degree
of freedom for the trap vibration. The displacement ${\bf x}$ of the particle
relative to the trap center ${\bf r}$ is chosen along the unit vector 
${\bf n}$ and written in terms of a creation operator $b$. 
The interaction potential reads
\begin{equation}
V( {\bf r}, t ) 
=
- {\bf x} \cdot {\bf F}( {\bf r}, t ) 
=
- a \left( b + b^\dag \right) {\bf n} \cdot {\bf F}( {\bf r}, t ) 
,
\label{eq:interaction}
\end{equation}
where $a = (\hbar /(2 M \Omega))^{1/2}$ 
is the size of the trap ground state 
($M$ is the particle mass and $\Omega$ the trap frequency)
and ${\bf F}( {\bf r}, t )$
the force acting on the particle. This force is fluctuating, and 
it is convenient to use a reduced density matrix description 
for the particle when the force fluctuations are averaged over.
The density matrix $\rho$ evolves according to a master equation that is
written in eq.(\ref{eqm:master2}) of appendix~\ref{a:master} for
a general coupling. For the Hamiltonian~(\ref{eq:interaction}), 
we get the following relaxation dynamics \cite{Gardiner} 
\begin{eqnarray}
\dot{\rho}|_{\rm relax} & = & - \frac{ \gamma_+ }{ 2 }
\left(
b^\dag b \rho + \rho b^\dag b - 2 b \rho b^\dag \right)
\nonumber\\
&& - \frac{ \gamma_- }{ 2 }
\left(
b b^\dag \rho + \rho b b^\dag - 2 b^\dag \rho b \right)
.
\label{eq:master}
\end{eqnarray}
In this equation, the transition rates 
$\gamma_\pm = \gamma( {\bf r}; \pm \Omega )$
are proportional to the spectral density $S_F^{ij}$ of the
force fluctuations taken at the trap vibration frequency $\Omega$
\begin{equation}
\gamma( {\bf r}; \omega ) =
\frac{ a^2 }{ \hbar^2 }
\sum_{ij}
n_i n_j
S_F^{ij}( {\bf r}; \omega )
,
\label{eq:def-rate}
\end{equation}
where $S_F^{ij}( {\bf r}; \omega )$ is defined by
\begin{equation}
S_F^{ij}( {\bf r}; \omega ) = 
\int\limits_{-\infty}^{+\infty}\!
d\tau
\left\langle F_i( {\bf r}, t + \tau ) F_j( {\bf r}, t ) \right\rangle
\,e^{ i\omega \tau }
.
\label{eqm:def-spectral-density}
\end{equation}

From the master equation~(\ref{eq:master}), it is easy to obtain
rate equations for the populations of the trap levels. For the ground state
population $\rho_{00} = \langle 0 | \rho | 0 \rangle$,
we get
\begin{equation}
\left.
\dot{\rho}_{00}
\right|_{\rm relax}
=
- \gamma_- \rho_{00} + \gamma_+ \rho_{11} 
.
\label{eq:rate-equation}
\end{equation}
Note that the transitions towards higher (lower) trap levels
occur with a rate equal to $\gamma_-$ (to $\gamma_+$).
In particular, the quantity $\gamma_-$ gives the depletion rate 
of the ground state population. The heating rate we are interested
in thus equals
\begin{equation}
\Gamma_{0\to1}( {\bf r} ) = 
\gamma_- = 
\frac{ a^2
}{ \hbar^2 }
\sum_{ij}
n_i n_j
S_F^{ij}( {\bf r}; - \Omega )
.
\label{eq:heating-rate}
\end{equation}
Note that the same result may be obtained from Fermi's Golden Rule,
by assuming a mixture of initial states for the fluctuating force field 
and summing over its final states. In Secs.~\ref{s:ion}
and \ref{s:spin}, the heating rates for trapped ions and spins are
computed using~(\ref{eq:heating-rate}). The main goal of the calculation   
is therefore the spectral density of the relevant force 
(electric or magnetic fields).

Finally, the master equation~(\ref{eq:master}) also allows to describe
the decay of the coherences between trap states which is a hazardeous 
process for quantum bit manipulations. The coherence between the lowest
trap levels relaxes according to 
\begin{equation}
\left.
\dot{\rho}_{01}
\right|_{\rm relax}
=
- \frac{ \gamma_+ + \gamma_- }{ 2 } \rho_{01} 
+ \sqrt{2} \gamma_+ \rho_{12} 
.
\label{eq:decoherence}
\end{equation}
We see that the coherences decay with a similar rate as the populations.
This is a consequence of the interaction Hamiltonian~(\ref{eq:interaction}),
and different results are obtained using other couplings or adding
explicit phase noise, see, e.g., Refs.\cite{Milburn98,Knight98}.
In the following, we focus on the population dynamics for simplicity.

\subsection{Internal state flips}
\label{s:loss}

In magnetic or optical traps for neutral particles, the trap potential 
depends on the internal atomic state (see fig.\ref{fig:model}, right
part). If this state is changed due to fluctuations in the magnetic field,
the particle may be subject to an anti-trapping potential and
strongly perturbed.
The interaction Hamiltonian for spin flips $|i\rangle \to |f\rangle$
is the Zeeman interaction
\begin{equation}
V_Z( {\bf r}, t ) = - \mbox{\boldmath$\mu$} \cdot {\bf B}( {\bf r}, t )
,
\label{eq:Zeeman-interaction}
\end{equation}
where {\boldmath$\mu$} is the particle's magnetic moment and 
${\bf B}( {\bf r}, t )$ the fluctuating part of the magnetic field.
For this interaction, a master equation similar to~(\ref{eq:master}) may
be formulated from the general theory outlined in appendix~\ref{a:master}.
This equation is not very instructive, however, if we assume that
the particle is lost as soon as it reaches the state $|f\rangle$.
In this case, it is sufficient to quote the transition rate 
$\Gamma_{i\to f}$ obtained from~(\ref{eqm:master2})
\begin{equation}
\Gamma_{i\to f}( {\bf r} )
=
\sum_{\alpha\beta}
\frac{ 
\langle i | \mu_\alpha | f \rangle \,
\langle f | \mu_\beta | i \rangle }{ \hbar^2 }
S_B^{\alpha\beta}( {\bf r}; -\omega_{fi} )
,
\label{eq:loss-rate}
\end{equation}
where $S_B^{\alpha\beta}$ is the magnetic field fluctuation spectrum defined
by an expression similar to~(\ref{eqm:def-spectral-density}),
and $\hbar \omega_{fi} = E_f - E_i$ the energy difference between
initial and final internal states. 
(We switch to greek subscripts to avoid confusion
with the initial state label.) In a magnetic trap, e.g.,
$|i\rangle$, $|f\rangle$ 
are magnetic sublevels and the frequency $\omega_{fi}$ a Larmor frequency
in the bias field of the trap.
In optical traps, we consider the hyperfine components of the atomic
ground state, $\omega_{fi}$ is thus the hyperfine splitting.


\section{Heating of a trapped charge}
\label{s:ion}

In this section, the master equation of the previous section is applied
to the most simple situation, that of an electrically charged particle
in a harmonic trap \cite{Wineland75,Lamoreaux97,James98,%
Milburn98,Knight98,Wineland98}. 
As mentioned in the introduction, the ion is heated up
because fluctuating electric fields leak out of the metallic
surface nearby. 
The force in the interaction Hamiltonian~(\ref{eq:interaction}) is
given by the electric field
\begin{equation}
{\bf F}( {\bf r}, t) = q {\bf E}( {\bf r}, t )  
\label{eq:ion-coupling}
\end{equation}
where $q$ is the ion's charge and ${\bf r}$ the position of the trap center.

\subsection{Electric field fluctuations}

In the formula~(\ref{eq:heating-rate}) for the heating rate, 
we need the spectral
density of the electric field fluctuations $S_E^{ij}( {\bf r}; \omega )$.
This quantity is conveniently obtained by making use of the
fluctuation-dissipation theorem outlined in appendix~\ref{a:FD}.
According to this theorem, the field's spectral density is proportional
to the imaginary part of the field's Green function 
$G_{ij}( {\bf r}, {\bf r}; \omega )$, multiplied
with the Bose-Einstein mean occupation number (eq.(\ref{eqa:FD})).
The geometry we have chosen is sufficiently simply to allow
the Green function to be calculated analytically \cite{Agarwal75a}. 
Recall that 
the Green function describes the electric field radiated by
an oscillating dipole (cf.\ eq.(\ref{eqm:def-Green})). This field
is the sum of the dipole field in free space plus the field reflected
from the surface. The free space field leads to a term 
$G^{(bb)}_{ij}( {\bf r}, {\bf r}; \omega )$  in the
Green function that is actually
independent of the trap position ${\bf r}$; it gives the spectral
density of the blackbody field (the Planck law)
\begin{eqnarray}
S_E^{(bb)ij}( {\bf r}; \omega ) & = &
S_E^{(bb)}( \omega ) \delta_{ij}
,
\label{eq:blackbody-spectrum}
\\
S_E^{(bb)}( \omega ) & = &
\frac{ \hbar \omega^3 
}{ 3\pi\varepsilon_0 c^3 ( 1 - e^{-\hbar\omega / T } )}
\label{eq:Planck-law}
\end{eqnarray}
where $T$ is the temperature of the surface (we put the Boltzmann 
constant $k_B = 1$).

To calculate the field reflected from the surface, we expand the
free space dipole field in plane waves and apply the Fresnel
reflection coefficients $r_{s,p}( u )$ 
for each wave incident on the surface (s and p label the two transverse
field polarizations and $u$ is the sine of the angle of incidence).
The resulting Green function 
$G^{(nf)}_{ij}( {\bf r}, {\bf r}; \omega )$ characterizes
the modification of the thermal radiation 
in the near field of the surface. The radiation density is increased
with respect to the far field expression~(\ref{eq:blackbody-spectrum})
because it also contains non-propagating (evanescent) waves. 
The corresponding spectral density depends only on the distance $z$
to the surface and may be written in the form
\cite{Agarwal75a}
\begin{equation}
S_E^{(nf)ij}( {\bf r}; \omega ) =
S_E^{(bb)}( \omega ) g_{ij}( k z )
\label{eq:prox-E-field}
\end{equation}
where the diagonal tensor $g_{ij}$ has the dimensionless elements
$g_{xx} = g_{yy} = g_\Vert$ and $g_{zz} = g_\perp$ with ($k = |\omega| / c$)
\begin{eqnarray}
g_\Vert( k z ) & = & \frac34 \mathop{\rm Re}\, \int\limits_0^{+\infty}\!\frac{ u\, du }{ v }
e^{2 i k z v }
\left( r_s(u) + (u^2 - 1 ) r_p( u ) \right) 
,
\nonumber
\\
g_\perp( k z ) & = & \frac32 \mathop{\rm Re}\, \int\limits_0^{+\infty}\!\frac{ u^3\, du}{ v}
e^{2 i k z v } r_p(u)
,
\\
v & = & \left\{
\begin{array}{ll}
\sqrt{ 1 - u^2 }, &\quad 0 \le u \le 1
,
\\
i \sqrt{ u^2 - 1}, &\quad u \ge 1
.
\end{array}
\right. 
\label{eq:g-tensor}
\end{eqnarray}
Finally, the relevant Fresnel coefficients are  
\begin{eqnarray}
r_p( u ) & = &
\frac{ \varepsilon v - \sqrt{ \varepsilon - u^2 } }{
\varepsilon v + \sqrt{ \varepsilon - u^2 } }
,
\nonumber
\\
r_s( u ) & = &
\frac{ v - \sqrt{ \varepsilon - u^2 } }{
v + \sqrt{ \varepsilon - u^2 } }
\label{eq:Fresnel-rsp}
\end{eqnarray}
where $\varepsilon(\omega)$ is the relative dielectric function of the bulk
metal. 

For typical trap frequencies the corresponding electromagnetic
wavelength is much larger than $z$, so we can restrict our calculations 
to the quasi-static limit $z \ll \lambdabar$ and find analytical expressions
for the tensor elements~(\ref{eq:g-tensor}). The details are outlined
in appendix~\ref{a:quasi-static-limit}. We have to distinguish between
the case of a large and a small skin depth of the conducting 
material compared to the distance $z$. The skin depth, which is the 
characteristic length scale on which an electromagnetic wave 
entering a conducting solid is damped, is given by (for
$\omega > 0$) \cite{Jackson}   
\begin{equation}
\delta
= \frac{1}{k}\sqrt{2 \varepsilon_0 \varrho \omega}
\end{equation}
where $\varrho$ is the specific resistance.
Since in our frequency regime the dielectric function for a metal is dominated 
by the zero--frequency pole, it is related to the skin depth by 
\begin{equation}
\varepsilon(\omega) 
\approx \frac{i}{\varepsilon_0 \varrho \omega } 
= \frac{2i}{k^2 \delta^2 }
.
\label{eq:eps-and-delta}
\end{equation}
In appendix~\ref{a:E-quasi-static-limit}, we derive approximations
for the functions $g_{\Vert,\perp}( k z )$ in the form of inverse power laws
(eqs.(\ref{eq:g-short-asymptotics},\ref{eq:g-intermediate-asymptotics})).
Both regimes of large and small skin depth can be covered by the following 
interpolation formula
\begin{equation}
g_{ij}( k z ) =
\frac{ 3 \delta^2 }{ 8 k z^3 }
\left(
s_{ij} + \delta_{ij} \frac{ z }{ \delta }
\right)
\label{eq:result-g-tensor}
\end{equation}
where $s_{ij}$ is a diagonal tensor with the elements 
$s_{xx} = s_{yy} = \frac12$, $s_{zz} = 1$.
Thus we arrive at a final expression for the electric field spectrum,
applying the high temperature limit of the Planck law~(\ref{eq:Planck-law}): 
\begin{equation}
S_E^{(nf)ij}( {\bf r}; \omega ) = 
\frac{T\varrho}{4\pi z^3}
\left( 
s_{ij} + \delta_{ij} \frac{z}{\delta( |\omega| )}
\right)
.
\label{eq:prox-E-field-2}
\end{equation}
We note that in the case of a short distance, the parallel and perpendicular 
tensor elements both show a $1/z^3$-dependence and differ by a factor of 2, 
whereas for larger distances the tensor elements are 
equal and show a $1/z^2$-behavior.

The $1/z^3$ power law of the
regime $z \ll \delta$ may be understood in terms of image theory:
the electrostatic dipole field varies precisely as $1/r^3$ and its 
reflection from the surface is characterized by the factor 
$(\varepsilon - 1)/(\varepsilon + 1) \approx 1 + i (k\delta)^2$.
The imaginary part of the reflected field thus 
reproduces~(\ref{eq:result-g-tensor}). This is the regime discussed
in Ref.\cite{Henkel99b}. It is interesting to note that
for a larger distance $z \gg \delta$, the field fluctuations are 
enhanced with respect to the electrostatic regime 
(see fig.~\ref{fig:ion-heating}). This is due to
the fact that the dipole field is more efficiently damped in the conductor
because the exponential decay in the skin layer quenches the
algebraic penetration of the field.

For completeness, we also mention the limiting case of a perfectly 
conducting surface 
($\varepsilon \to \infty$) whose skin depth $\delta$ 
vanishes. The previous asymptotic expansion does not cover this
case. The coefficients $g_{\Vert,\perp}( k z )$ given
in the appendix~\ref{a:quasi-static-limit}, eq.(\ref{eq:pc-asymptotics}),
show
damped oscillations with a period equal to the wavelength. In the
short-distance limit $z \ll \lambdabar$, we get
$g_\Vert( kz ) \to - 1$ and $g_\perp( kz ) \to 1$, the divergence
at $z \to 0$ thus disappears. The electric field fluctuations
are essentially those of the free space blackbody spectrum, with a minor 
modification due to the boundary conditions.

\subsection{Heating rate}

\begin{figure}[tbh]
\resizebox{\columnwidth}{!}{%
\includegraphics*{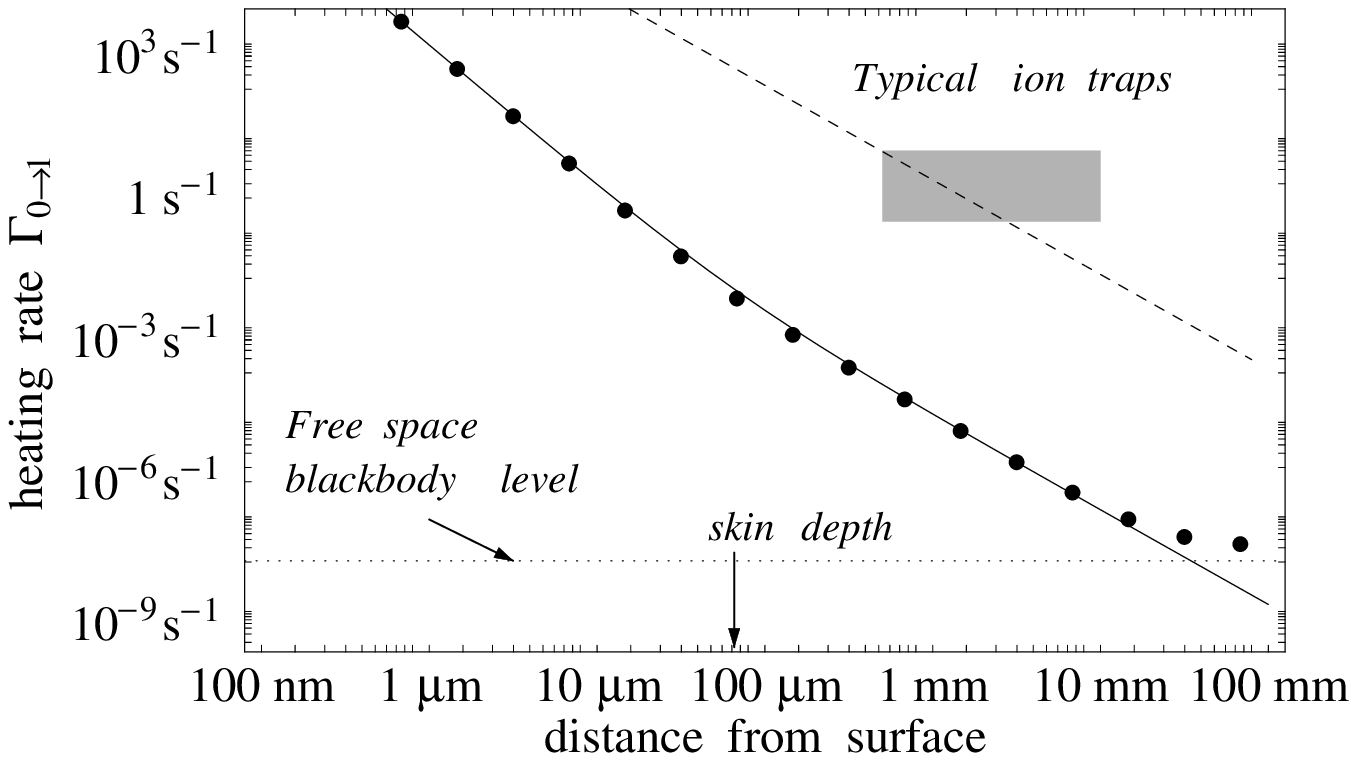}
}
\caption[]{\label{fig:ion-heating}%
Heating rate for a trapped ion.
Dots: coupling to electric proximity fields,
computed from~(\ref{eq:g-tensor}). The solid line is obtained using
the asymptotic formula~(\ref{eq:result-g-tensor}).
Dashed line: coupling to thermal voltage fluctuations.
\newline
Parameters: trap frequency $\Omega/2\pi = 1\,$MHz, 
copper substrate with $\varrho = 1.7\times10^{-6}\,{\rm \Omega\,cm}$
at $T = 300\,{\rm K}$. The ion mass is $M = 40\,{\rm amu}$, 
and its charge $q = e$. The trap axis is perpendicular to the surface,
${\bf n} = {\bf e}_z$. The thermal voltage fluctuations 
are characterized by a circuit resistance $1\,{\rm\Omega}$ \cite{Lamoreaux97}. 
The endcaps are separated by twice the ion-surface distance.
Size and inverse lifetimes of typical ion traps are indicated
by the shaded rectangle \cite{Wineland96,Blatt99p,Peik99p}.
}
\end{figure}

We plot in fig.~\ref{fig:ion-heating} the heating rate~(\ref{eq:heating-rate})
for an ion (trap frequency $\Omega/2\pi= 1\,$MHz)
above a copper surface. 
The dots are based on an exact (numerical)
evaluation of the $g$-coefficients~(\ref{eq:g-tensor}), while the
solid line uses the interpolation~(\ref{eq:result-g-tensor}). 
The change in the power law at the skin depth is clearly visible.
Note the marked increase of the field fluctuations compared to the
free space blackbody level (dotted line). Also shown is the estimate given
by Lamoreaux~\cite{Lamoreaux97} who modeled the trap in terms of
a resistively damped capacitor with a thermally fluctuating voltage
(Johnson noise). Wineland et al.\ \cite{Wineland98} pointed out
that realistic estimates for the corresponding resistance actually give
smaller heating rates. Our results suggest that the miniaturization
of ion traps down to $\mu$m sizes entails difficulties to maintain 
long coherent storage times, unless all physical components are cooled
down.


\section{Trapped spin coupling to magnetic fields}
\label{s:spin}

In this section, we turn to traps for neutral particles
and consider the Zeeman coupling~(\ref{eq:Zeeman-interaction})
of the atomic magnetic moment to a fluctuating magnetic field.
In magnetic and optical traps, this coupling may induce a spin flip 
to a non-trapped state (magnetic sublevel or hyperfine state).
This implies a nonzero loss rate from the trap that we calculate
in subsection~\ref{s:spin-flips}. 
On the other hand, the Zeeman interaction also exerts a force 
proportional to the gradient of the magnetic field. If this force
fluctuates, it does not necessarily flip the atomic spin, but 
excites the atom into a higher trap level. The corresponding heating
rate is the subject of subsection~\ref{s:spin-heating}.

\subsection{Spin flips}
\label{s:spin-flips}

\subsubsection{Magnetic field correlations.}
\label{s:magnetic-correlations}

We first compute the magnetic field fluctuations in the vicinity of
the solid surface. By analogy to the ion case, we use the 
fluctuation-dissipation theorem~(\ref{eqa:FD}) and determine
the Green tensor for the magnetic field. In fact, the calculation 
is very similar to that for the electric field: starting from 
the field radiated in free space, we expand it in spatial Fourier
components and compute for each plane wave the reflection at the
solid surface. 
It turns out that the Fresnel coefficients for the magnetic field 
are identical to those for electric fields,
except that one has to exchange the s- and p-polarizations. We thus
get the following near-field correction to the 
magnetic field fluctuation spectrum
\begin{equation}
S_B^{(nf)ij}( {\bf r}; \omega ) =
\frac{ S_E^{(bb)}( \omega ) }{ c^2 }
h_{ij}\!\left( k z \right)
.
\label{eq:prox-B-field}
\end{equation}
Similar to~(\ref{eq:prox-E-field}), 
$h_{ij}$ is a dimensionless and diagonal tensor with elements
\begin{eqnarray}
h_\Vert( k z ) & = & \frac34 \mathop{\rm Re}\, \int\limits_0^{+\infty}\!\frac{ u\, du }{ v }
e^{2 i k z v }
\left( r_p(u) + (u^2 - 1 ) r_s( u ) \right) 
,
\nonumber
\\
h_\perp( k z ) & = & \frac32 \mathop{\rm Re}\, \int\limits_0^{+\infty}\!\frac{ u^3\, du}{ v}
e^{2 i k z v } r_s(u)
.
\label{eq:def-h-elements}
\end{eqnarray}
For experimentally relevant parameters, the magnetic fields at the
resonance frequency have a wavelength (at least some cm) much longer
than the size of the trap. This implies again that we need the short-distance
asymptotics $z \ll \lambdabar$ of~(\ref{eq:prox-B-field}). 
A calculation outlined in appendix~\ref{a:B-quasi-static-limit}
gives the following interpolation formula that covers both regimes of
a large and small skin depth
\begin{equation}
h_{ij}( k z ) = \frac{ 3 s_{ij} }{ 8 k^3 \delta^2 z }
\left( 1 + \frac{ 2 z^3 }{ 3 \delta^3 } \right)^{-1}
\label{eq:result-h-tensor}
\end{equation}
where $s_{ij}$ is the diagonal tensor introduced in~(\ref{eq:result-g-tensor}).
The magnetic field spectrum~(\ref{eq:prox-B-field}) thus equals 
in the high-temperature limit 
\begin{equation}
S_B^{(nf)ij}( {\bf r}; \omega ) =
\frac{ \mu_0^2 T }{ 16 \pi \varrho }
\frac{ s_{ij} }{ z }
\left( 1 + \frac{ 2 z^3 }{ 3 \delta^3( |\omega| ) } \right)^{-1}
.
\label{eq:prox-B-field-2}
\end{equation}
Note the different exponents for the distance dependence compared to
the electric field fluctuations~(\ref{eq:prox-E-field-2}).

If the trap distance is small compared to the skin depth, we recover
the magnetic field spectrum given in eq.(10) of \cite{Henkel99b},
apart from the fact that the parallel tensor components 
($s_{xx}$, $s_{yy}$)
differ. This difference is due to the fact that the calculation of
\cite{Henkel99b} uses the Biot-Savart law to get the magnetic field
from a statistical model of polarization currents in the solid. 
This approach is valid for stationary currents only, and a difficulty 
appears at the surface
because the model for the currents is not divergence-free there. Therefore,
while the magnetic field perpendicular to the surface is correctly
described, the parallel components are overestimated.


\subsubsection{Internal matrix elements.}

In order to compute the spin flip loss rate we have to evaluate 
matrix elements of the total magnetic moment operator as indicated 
in~(\ref{eq:loss-rate}). This operator is in general given by 
\begin{equation}
{\bf \mu} = -\mu_B \left( g_L {\bf L} + g_S {\bf S} 
- g_I \frac{m_e}{m_p} {\bf I}\right),
\label{eq:total-magnetic-moment}
\end{equation}
with $\mu_B$ the Bohr magneton, ${\bf L}$ the total orbital angular momentum operator, 
${\bf S}$ the electronic spin operator, ${\bf I}$ the nuclear spin operator and 
$g_L$, $g_S$ and $g_I$ the corresponding $g$-factors. 
Since the proton mass $m_p$ is larger than the electron mass $m_e$ by three orders 
of magnitude, we can neglect the contribution of the nuclear magnetic moment.
Furthermore, the reasonable restriction to an atomic ground state with $L=0$ 
reduces the problem to the calculation of matrix elements of solely the spin 
operator. Together with the fact that the tensor $h_{ij}$ 
in~(\ref{eq:result-h-tensor}) for the magnetic field correlations is diagonal, 
we can focus on terms of the form
\begin{equation}
\left|\left\langle f \right|\mu_{\alpha} \left|i\right\rangle\right|^2 
=\mu_{B}^2 g_{S}^2\left|\left\langle f \right|
S_{\alpha}\left|i\right\rangle\right|^2.
\label{eq:spin-moment-probability}
\end{equation}
In the following we will restrict ourselves to two extreme cases:
the coupling between two Zeeman sublevels in the 
presence of an external magnetic field and the coupling between two hyperfine 
ground states without external fields applied. 
The former case is e.g. realized in a magnetic trap, 
whereas the latter corresponds to optical traps.  

In the case of a magnetic trap the trapped atom is subject to a constant 
magnetic field with strength $B_0$ in the center of the trap, assuming the atom
is not moving. 
The magnetic sublevels are split due to the Zeeman effect by the Larmor frequency 
$\omega_L=g_S\mu_B B_0/\hbar$. 
(We focus on a vanishing nuclear spin for simplicity.)
Without loss of generality we can assume the magnetic 
field to be lying within the $xz$-plane, since the diagonal tensor in 
(\ref{eq:result-h-tensor}) has the symmetry property $h_{xx}=h_{yy}$.
If the magnetic field forms an angle $\theta$ with respect to the $z$-axis, we 
denote by $|m\rangle_\theta$ the basis states with quantization axis
parallel to the magnetic field (the `trap basis'). 
Rewriting~(\ref{eq:spin-moment-probability})
leaves us to calculate matrix elements of the form
\begin{equation} 
\left|\left\langle f \right|\mu_{\alpha}\left| i \right\rangle\right|^2
=\mu_B^2 g_S^2\left|{}_{\theta}\langle m_f | 
S_{\alpha}| m_i \rangle_{\theta}\right|^2.
\label{eq:zeeman-probability}
\end{equation}
These elements are evaluated by expanding the spin vector components
in a rotated coordinate system (denoted by the prime) adapted to the trap basis.
The result is the following:
\begin{eqnarray}
{}_{\theta}\langle m_f | S_x | m_i \rangle_{\theta}
&=& \left({}_{\theta}\langle m_f | S_+^{\prime} | m_i \rangle_{\theta} + 
    {}_{\theta}\langle m_f | S_-^{\prime} | m_i \rangle_{\theta}\right)
    \frac{\cos{\theta}}{2} 
    \nonumber \\
&+& {}_{\theta}\langle m_f | S_3^{\prime} | m_i \rangle_{\theta}
    \sin{\theta},                 
    \nonumber \\
{}_{\theta}\langle m_f | S_y | m_i \rangle_{\theta}
&=& \frac{i}{2}\left({}_{\theta}\langle m_f | 
    S_-^{\prime} | m_i \rangle_{\theta} - 
    {}_{\theta}\langle m_f | S_+^{\prime} | m_i \rangle_{\theta} 
    \right),                      
    \nonumber \\
{}_{\theta}\langle m_f | S_3 | m_i \rangle_{\theta}
&=& \left({}_{\theta}\langle m_f | S_+^{\prime} | m_i \rangle_{\theta} +
    {}_{\theta}\langle m_f | S_-^{\prime} | m_i \rangle_{\theta}\right)
    \frac{ {-\sin{\theta}} }{ 2 }     
    \nonumber \\
&+& {}_{\theta}\langle m_f | S_3^{\prime} | m_i \rangle_{\theta}\cos{\theta}
\label{eq:zeeman-matrix-elements}
\end{eqnarray}
where $S_3^{\prime}$ is the $z$-component of the spin operator and 
$S_+^{\prime}$, $S_-^{\prime}$ 
correspond to raising resp.\ lowering operators in the trap basis, whose action 
is known~\cite{Sakurai}. In the case of an electronic spin $S=1/2$, 
the trapped (untrapped) level is the $|m_i\rangle_\theta = |{-1/2}\rangle_\theta$ 
($|m_f\rangle_\theta = |1/2 \rangle_\theta$) Zeeman sublevel,
respectively. The matrix elements~(\ref{eq:zeeman-matrix-elements}) then become
\begin{eqnarray}
{}_{\theta}\langle 1/2 | S_x | {-1/2} \rangle_{\theta}
&=& \frac{\cos{\theta}}{2},         \nonumber \\
{}_{\theta}\langle 1/2 | S_y | {-1/2} \rangle_{\theta}
&=& -\frac{i}{2},                   \nonumber \\
{}_{\theta}\langle 1/2 | S_z | {-1/2} \rangle_{\theta}
&=& -\frac{\sin{\theta}}{2}. 
\label{eq:zeeman-matrix-elements-special}
\end{eqnarray}
With this result, we can compute the magnetic loss rate~(\ref{eq:Larmor-loss})
below. 

In the case of an optical trap we have to take into account that the
nuclear spin couples to the electronic spin, 
${\bf F}={\bf S}+{\bf I}$, and causes the ground state to split into 
hyperfine levels, separated by a frequency $\omega_{HF}$. 
We are now
interested in the transition probability from one hyperfine ground
state to another. Thus, for this case we can 
write~(\ref{eq:spin-moment-probability}) as 
\begin{equation}
\left|\left\langle f \right| \mu_{\alpha} \left| i \right\rangle\right|^2
=\mu_B^2 g_S^2\left|\left\langle F_f \right| 
  S_{\alpha} \left| F_i \right\rangle\right|^2
\label{eq:hyperfine-probability}
\end{equation}
A transition from one hyperfine ground state to another can take place 
between different magnetic sublevels. Thus we first have to calculate
the transition rate between two of these states. This is done by
expanding the basis states in the uncoupled basis,
choosing the quantization axis taken along the $z$-axis:
\begin{equation}
\left| F m \right\rangle = \sum_{m_S, m_I} C^{m_S{ }m_I}_{F{ }m}
\left| m_S m_I \right\rangle
\label{eq:basis-expansion}
\end{equation}
where $C^{m_S{ }m_I}_{F{ }m}$ are the Clebsch--Gordan
coefficients. The matrix element between two hyperfine magnetic levels 
is then
\begin{eqnarray}
&& \left\langle F_f m_f \right| S_{\alpha} \left| F_i m_i \right\rangle
\nonumber\\
&& = \sum_{m_S, m_S',
  m_I}C^{m_S'{ }m_I}_{F_f{ }m_f}C^{m_S{ }m_I}_{F_i{ }m_i}
\left\langle m_S' \right| S_{\alpha} \left| m_S \right\rangle.
\label{eq:transition-hyperfine-zeeman}
\end{eqnarray}
Note that the nuclear spin does not flip in the transition.
Again the action of $S_{\alpha}$ 
onto the electronic spin states $|m_S\rangle$ is well--known
in~(\ref{eq:transition-hyperfine-zeeman}).
We obtain an effective transition rate between the two
hyperfine manifolds by summing the rates  
over all final $m_f$-levels and taking the average over the initial
$m_i$-levels. This gives the following result for 
the hyperfine matrix element~(\ref{eq:hyperfine-probability})
\begin{equation}
\left|\left\langle F_f \right| S_{\alpha} \left| F_i \right\rangle\right|^2 
= \frac{1}{2F_i+1} \sum_{m_f, m_i} 
\left|\left\langle F_f m_f \right| S_{\alpha} \left| F_i m_i \right\rangle\right|^2.
\label{eq:averaged-hyperfine-probability}
\end{equation}
We finally note that this calculation assumes that the frequencies for the
transitions $|F_i m_i\rangle \to |F_f m_f\rangle$ are all equal to the
hyperfine splitting $\omega_{HF}$. This is a good approximation if 
$\hbar\omega_{HF}$ is large compared to the optical trap potential (that may
lift the degeneracy of the hyperfine states even without a static magnetic field).


\subsubsection{Loss rate.}

\begin{figure}[tbh]
\resizebox{\columnwidth}{!}{%
\includegraphics*{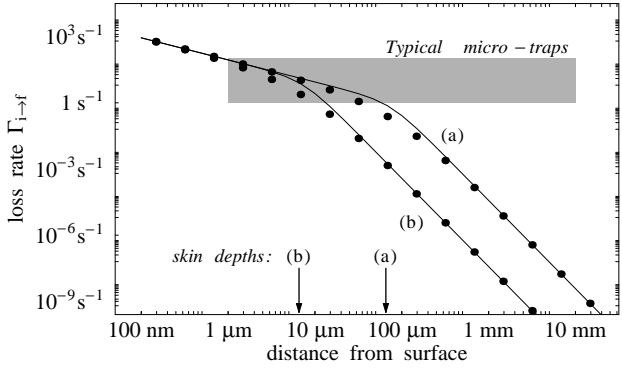}
}
\caption[]{\label{fig:flip-rates}%
Loss rates in a magnetic trap above a copper surface. 
Dots (solid lines): results
based on~(\ref{eq:def-h-elements}) (on the asymptotic 
interpolation~(\ref{eq:result-h-tensor})). Results for two different
Larmor frequencies $\omega_L/2\pi = 1\,$MHz (curve a)
and $100\,$MHz (curve b) are shown.
The arrows mark the corresponding skin depths. The shaded area
indicates experimental data obtained in Konstanz and Heidelberg
\cite{Ovchinnikov97b,Mlynek98b}.
\newline
Parameters: spin $S=1/2$, magnetic bias field aligned parallel 
to the surface. The loss rate due to the blackbody field (the prefactor
in~(\ref{eq:Larmor-loss-special})) is about 
$10^{-13}\,{\rm s}^{-1}$ at $100\,$MHz (not shown).
}
\end{figure}

Combining the matrix elements~(\ref{eq:spin-moment-probability})
for the magnetic moment, the magnetic field 
spectrum~(\ref{eq:prox-B-field-2}) and eq.(\ref{eq:loss-rate}), 
we get the following loss rate for a magnetic trap 
\begin{equation}
\Gamma_{i\to f}( {\bf r} )
= \frac{ \mu_B^2 g_S^2 \omega_L^2 T }{ 3\pi\varepsilon_0 
\hbar^2 c^5 }
\sum_{\alpha}
(h_{\alpha\alpha}( k z ) + 1)
\left| \langle f | S_\alpha | i \rangle \right|^2
.
\label{eq:Larmor-loss}
\end{equation}
For the case of an electronic spin $S=1/2$ and no nuclear spin we can 
use the matrix elements from~(\ref{eq:zeeman-matrix-elements-special}) 
and obtain
\begin{eqnarray}
\Gamma_{-\frac{1}{2}\to\frac{1}{2}}( {\bf r} ) 
& = & 
\frac{ \mu_B^2 g_S^2 \omega_L^2 T }{ 12\pi\varepsilon_0 
\hbar^2 c^5 }
\left\{(h_{\parallel}(kz)+1)(1+\cos^2{\theta}) + {}
\right.
\nonumber\\
&& \qquad
\left.
{}+(h_{\perp}(kz)+1)\sin^2{\theta}\right\}
.
\label{eq:Larmor-loss-special}
\end{eqnarray}
This loss rate is plotted in fig.~\ref{fig:flip-rates} for 
two different Larmor frequencies $\omega_L$, with the 
trap bias field chosen parallel to the surface ($\theta = \pi/2$).
We see that quite large loss rates occur if the trap center approaches 
the surface down to a few micrometers. Again, miniaturized traps have
to face the influence of larger noise fields. 

\begin{figure}[tbh]
\resizebox{\columnwidth}{!}{%
\includegraphics*{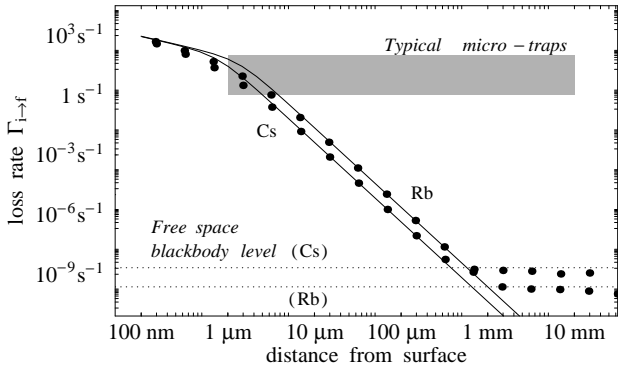}
}
\caption[]{\label{fig:HF-rates}%
Loss rates due to hyperfine-changing transitions
in an optical trap above a copper surface. 
Dots (solid lines): results
based on~(\ref{eq:def-h-elements}) (on the asymptotic 
interpolation~(\ref{eq:result-h-tensor})). Results for two different
atoms are shown: $^{85}$Rb ($I = 5/2$, $\omega_{HF}/2\pi =
3.04\,$GHz, transition $F_i = 2 \to 3 = F_f$) 
and $^{133}$Cs ($I = 7/2$, $\omega_{HF}/2\pi =
9.193\,$GHz, transition $F_i = 3 \to 4 = F_f$).
The horizontal dotted line marks the corresponding loss rates in the
free space blackbody field.
}
\end{figure}

In fig.~\ref{fig:HF-rates}, we plot the loss rates obtained 
from the effective matrix element~(\ref{eq:averaged-hyperfine-probability}) 
for hyperfine-changing transitions.
The data are calculated for the lower ground states of 
trapped $^{85}$Rb and $^{133}$Cs. One observes that
these rates are much smaller than those for magnetic traps. It is
interesting that this reduction is due to the skin effect:
indeed, the magnetic field fluctuations~(\ref{eq:prox-B-field-2})
in the intermediate-distance regime $\delta \ll z \ll \lambdabar$
are proportional to $\delta^{3} \propto \omega^{-3/2}$.
Larger transition frequencies thus lead to smaller loss rates.

\subsection{Heating of the c.m.\ motion}
\label{s:spin-heating}

This case is treated by analogy to the trapped ion. The 
Zeeman interaction~(\ref{eq:Zeeman-interaction}) gives the following 
magnetic force
\begin{equation}
{\bf F}_Z( {\bf r}, t ) =
\nabla \left(\mbox{\boldmath$\mu$} \cdot {\bf B}( {\bf r}, t ) \right)
\label{eq:Zeeman-force}
\end{equation}
that couples to the displacement 
of the particle from its equilibrium position. 
The matrix elements for the displacement are that of a 
1D harmonic oscillator and are given in subsection~\ref{s:heating}.
We are left with the calculation of the magnetic force's
spectral density. To this end, recall the identity
\begin{eqnarray}
&& \langle F_{Zi}( {\bf r}, t' ) F_{Zj}( {\bf r}, t ) \rangle
\nonumber\\
&& =
\left.
\frac{ \partial }{ \partial r_{1i} } 
\frac{ \partial }{ \partial r_{2j} }
\langle V_Z( {\bf r}_1, t' ) V_Z( {\bf r}_2, t' ) \rangle
\right|_{ {\bf r}_1 = {\bf r}_2 = {\bf r} }
.
\label{eq:get-force-correlations}
\end{eqnarray}
The relevant information is thus contained in the cross correlation 
function for the magnetic field at two different positions ${\bf r}_{1,2}$.
From the fluctuation-dissipation theorem (appendix~\ref{a:FD}), this
correlation function is proportional to the Green function 
$H_{ij}( {\bf r}_1, {\bf r}_2; \omega )$ for the magnetic field. 
To simplify the calculation, we focus on a trap with an axis ${\bf n}$
perpendicular to the surface. According to~(\ref{eq:heating-rate}), we
then only need the $zz$-component of the force fluctuation tensor.
In the identity~(\ref{eq:get-force-correlations}), it is thus sufficient
to take two positions ${\bf r}_{1,2} =
({\bf R}, z_{1,2})$ that differ only in the vertical coordinate
(${\bf R} = (x, y)$ denotes the coordinates parallel to the surface). 
It may now be shown that the surface-dependent part 
$H^{(nf)}_{ij}( {\bf r}_1, {\bf r}_2; \omega )$ of the Green tensor 
depends only on the average distance $\bar{z} = (z_1 + z_2)/2$
and the lateral separation ${\bf R}_2 - {\bf R}_1$ \cite{Agarwal75a}. 
This is clear,
e.g., from image theory. Since ${\bf R}_1 = {\bf R}_2$ 
for our special case, we may write 
\begin{equation}
H^{(nf)}_{ij}( {\bf R}, z_1, {\bf R}, z_2;  \omega )
=
H^{(nf)}_{ij}( {\bf R}, \bar{z}, {\bf R}, \bar{z} ; \omega )
\label{eq:trick-33}
\end{equation}
where the right-hand side is the Green function taken at identical
positions that has been calculated in subsection~\ref{s:magnetic-correlations}. 

We now use the results~(\ref{eq:h-short-asymptotics}, 
\ref{eq:h-intermediate-asymptotics}) for the magnetic correlation
tensor (app.~\ref{a:B-quasi-static-limit}), 
write $z = (z_1 + z_2)/2$ and differentiate with respect to $z_{1,2}$. 
All told, both asymptotic regimes of small and large skin depth 
are described by the interpolation formula 
\begin{equation}
S_{F_Z}^{zz}( {\bf r}; \omega ) =
\frac{ \mu_0^2 T }{ 64 \pi \varrho }
\frac{ \langle i | \mbox{\boldmath$\mu$}^2 + \mu_3^2 | i \rangle }{ z^3 }
\left( 1 + \frac{ z^3 }{ 15 \delta^3 }
\right)^{-1}
.
\label{eq:Zeeman-force-correlations}
\end{equation}
This spectrum is already summed over all final Zeeman states, assuming
that all of them are trapped. The average for the magnetic moment is taken 
in the initial state. For an atom with $L = 0, S = 1/2$ in the ground state,
it equals $g_S^2\mu_B^2 \approx 4 \mu_B^2$ where $\mu_B$ is the Bohr magneton.

If the trap distance is small compared to the skin depth, we recover the
expression~(11) of~\cite{Henkel99b} for the heating rate 
\begin{equation}
\Gamma_{0\to1}( {\bf r} ) 
= 
\frac{ \mu_0^2 T \mu^2_B g_S^2 }{ 64 \pi \hbar \Omega M \varrho \, z^3 }
,
\label{eq:spin-heating}
\end{equation}
apart from different weights for the parallel and perpendicular spin
components. This is due to the different magnetic field correlation 
tensor~(\ref{eq:prox-B-field-2}) that has already been discussed above.

\begin{figure}[tbh]
\resizebox{\columnwidth}{!}{%
\includegraphics*{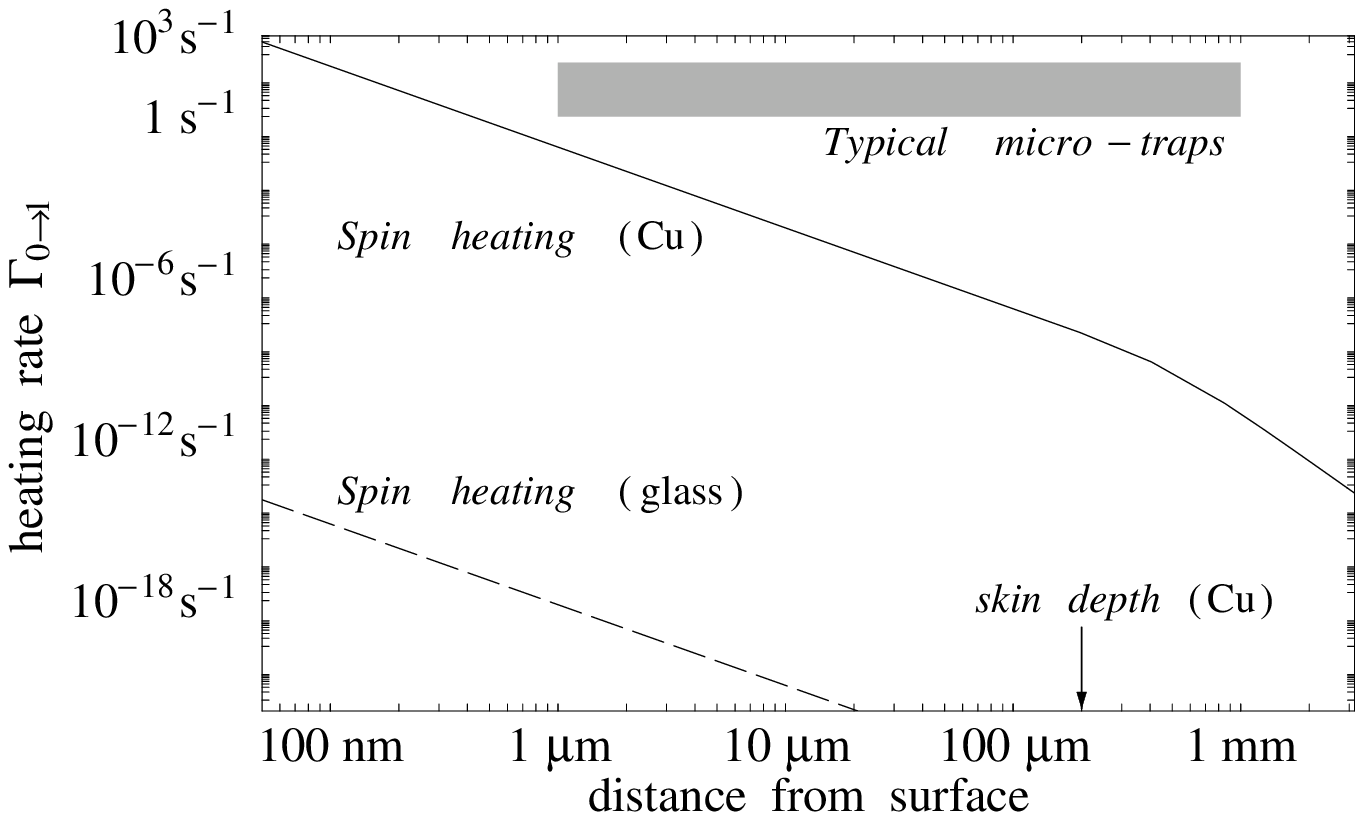}
}
\caption[]{\label{fig:spinHeating}%
Heating rate for a trapped spin above copper and glass substrates.
\\
Parameters: trap frequency $\Omega/2\pi = 100\,{\rm kHz}$,
$M = 40\,{\rm amu}$, magnetic moment $\mu = \mu_B = $ 1 Bohr magneton,
spin $S = 1/2$. The heating rate due to the magnetic blackbody field 
(not shown) is about $10^{-39}\, {\rm s}^{-1}$. For the glass substrate, a
dielectric constant with $\mathop{\rm Re}\,\varepsilon = 5$ and a specific
resistance $\varrho = 10^{11}{\rm \Omega}\,{\rm cm}$ are taken. These values
are used in the short-distance asymptotics~(\ref{eq:h-short-asymptotics})
to compute the magnetic field fluctuations. 
}
\end{figure}

In fig.~\ref{fig:spinHeating}, we plot the heating rate $\Gamma_{0\to1}$ 
obtained from the magnetic fluctuation 
spectrum~(\ref{eq:Zeeman-force-correlations}) for a typical trap above
both a copper and a glass surface. The heating rate above glass is much
smaller because glass is a poor conductor.
For a copper substrate, note the crossover when the distance becomes
larger than the skin depth. A remarkable result is the large value
of the heating rate for small traps (dimensions below the $\mu$m range).


\section{Summary and outlook}
\label{s:conclu}

To summarize, we have developed a theoretical framework for the
systematic investigation of the heating and concomitant loss of
coherence in small particle traps. Our results indicate a clear
predominance of near field effects over ordinary (free space)
blackbody radiation. They establish upper bounds for
life times in a variety of experimentally relevant types of traps.

The present model is restricted to particle motion in a single dimension,
and the extension to a three-dimensional trap geometry is an obvious
step for future work. A theory beyond the rate equations discussed
here could include noise-induced shifts of the particle's energy
levels. Finally, still other interactions might be considered for neutral 
atoms. The coupling to electric fields via the polarizability tensor 
is currently under investigation.

\begin{small}
\paragraph{Acknowledgments.}
C.~H. would like to thank R\'emi Carminati, Jean-Jacques Greffet, 
Karl Joulain, and Stefan Scheel for sharing their deep understanding of 
electromagnetic near-field spectra. 
We are indebted to John B. Pendry, Ekkehard Peik, and Ferdinand Schmidt-Kaler
for communicating results of previously unpublished work. 
Travel costs have been covered by Laboratoire d'Energ\'etique Mol\'eculaire 
et Macroscopique, Combustion of Ecole Centrale Paris, Ch\^atenay-Malabry, 
France.
This work has been supported by a research grant awarded to C.~H. 
by the Deutsche Forschungsgemeinschaft. 

\end{small}

\appendix


\section{Statistical tools}

\subsection{Master equations}
\label{a:master}

We outline here a general master equation \cite{Agarwal75a} 
that describes the reduced dynamics of a system coupled to a reservoir.
The coupling Hamiltonian is given in terms of an arbitrary system operator 
${\bf s}$, a fluctuating force ${\bf F}( {\bf r}, t)$, 
and a coupling constant $g$ 
\begin{equation}
V( {\bf r}, t ) = - g \, {\bf s} \cdot {\bf F}( {\bf r}, t)
.
\label{eq:system-bath-Hamiltonian}
\end{equation}
Throughout this paper, the parameter ${\bf r}$ denotes the trap center
position. For a trapped ion, e.g., the system operator ${\bf s}$ would 
describe the displacement of the ion from the trap center, 
see eq.(\ref{eq:interaction}).
In the Markov limit and ignoring reservoir-induced level shifts,
the relaxation dynamics of the reduced system density matrix $\rho$ 
is
\begin{eqnarray}
\dot{\rho}|_{\rm relax} & = & - \frac{ g^2 }{ \hbar^2 } \sum_{ij}
\frac{ S_F^{ij}( {\bf r}; \omega ) }{ 2 }
\left(
s_i^{(-)} s_j^{(+)} \rho + 
\rho s_i^{(-)} s_j^{(+)} 
\right.
\nonumber\\
&& \quad 
\left.
-
2 s_j^{(+)} \rho s_i^{(-)}
\right)
\nonumber\\   
&& -  \frac{ g^2 }{ \hbar^2 } \sum_{ij}
\frac{ S_F^{ij}( {\bf r}; -\omega ) }{ 2 }
\left(
s_i^{(+)} s_j^{(-)} \rho +
\rho s_i^{(+)} s_j^{(-)} 
\right.
\nonumber\\
&& \quad
\left.
-
2 s_j^{(-)} \rho s_i^{(+)} 
\right)
\label{eqm:master2}
\end{eqnarray}
where the ${\bf s}^{(\pm)}$ is the positive (negative) frequency part of
the system operator. More precisely, the free system evolution in the 
Heisenberg picture is given by
\begin{equation}
{\bf s}(t) = {\bf s}^{(+)} e^{-i \omega t}
+ {\bf s}^{(-)} e^{i \omega t}
\end{equation}
where $\hbar\omega \, (> 0)$ is the energy difference between two adjacent 
system states. The spectral density in~(\ref{eqm:master2}) is defined by
(cf.\ eq.(\ref{eqm:def-spectral-density}))
\begin{equation}
S_F^{ij}( {\bf r}; \omega ) = 
\int\limits_{-\infty}^{+\infty}\!
d\tau
\left\langle F_i( {\bf r}, t + \tau ) F_j( {\bf r}, t ) \right\rangle
\,e^{ i\omega \tau }
.
\label{eqa:def-spectral-density}
\end{equation}
The master equation~(\ref{eqm:master2}) allows to derive rate equations
similar to~(\ref{eq:rate-equation}), and these show 
that the rates proportional to $S_F^{ij}( {\bf r}; +\omega)$ govern 
spontaneous and stimulated decay processes, while excitation processes
are proportional to $S_F^{ij}( {\bf r}; -\omega)$. 
The latter correlation function is thus relevant for our heating problem.

\subsection{Fluctuation--dissipation theorem}
\label{a:FD}
 
In a reservoir at thermal equilibrium, there is a relation between
the cross correlation tensor for the field fluctuations 
and the field's Green tensor \cite{Agarwal75a}. This relation also
holds for correlations taken at different positions in space,
that we have to compute in subsection~\ref{s:spin-heating}.
For a force field ${\bf F}( {\bf r}, t )$,
the cross correlation tensor is defined by 
generalizing~(\ref{eqa:def-spectral-density})
\begin{equation}
S_F^{ij}( {\bf r}_1, {\bf r}_2; \omega )
= 
\int\limits_{-\infty}^{+\infty}\!
d\tau
\left\langle F_i( {\bf r}_1, t + \tau ) F_j( {\bf r}_2, t ) 
\right\rangle
\,e^{ i\omega \tau }
.
\label{eqm:def-cross-correlations}
\end{equation}
The Green function is defined as the force field created by a classical
monochromatic, localized disturbance ${\bf a}$ at ${\bf r}_0$ 
(e.g.\ the electric field of an oscillating point dipole).
The interaction Hamiltonian density is
\begin{equation}
- e^{-i \omega t} \delta( {\bf r} - {\bf r}_0 )
{\bf a}\cdot{\bf F}( {\bf r}, t )
.
\end{equation}
In thermal equilibrium, the average linear response to this source 
is a harmonic field $\langle {\bf F}( {\bf r}, t; {\bf r}_0 ) \rangle$
that depends parametrically on the source position ${\bf r}_0$ and is 
proportional to the displacement ${\bf a}$. 
The Green function is the corresponding proportionality factor
\begin{equation}
\langle F_i( {\bf r}, t; {\bf r}_0 ) \rangle 
= e^{-i \omega t}
\sum_j
G_{ij}( {\bf r}, {\bf r}_0 ; \omega ) a_j 
.
\label{eqm:def-Green}
\end{equation}
(The averaging $\langle \cdots \rangle$ removes
the oscillations of the free field.)
The fluctuation-dissipation theorem now states \cite{Agarwal75a}
\begin{equation}
S_F^{ij}( {\bf r}_1, {\bf r}_2; \omega )
=
\frac{ 2 \hbar }{ 1 - e^{- \hbar \omega / T} }
\mathop{\rm Im}\, G_{ij}( {\bf r}_1, {\bf r}_2; \omega )
.
\label{eqa:FD}
\end{equation}
Note that in terms of the mean thermal occupation number
$\bar{n}_{\rm th} = 1/(e^{\hbar\omega/T} - 1)$, one has
(for $\omega > 0$)
\begin{eqnarray}
S_F^{ij}( {\bf r}_1, {\bf r}_2; \omega )
& = & 
2 \hbar \left( \bar{n}_{\rm th} + 1 \right)
\mathop{\rm Im}\, G_{ij}( {\bf r}_1, {\bf r}_2; \omega )
,
\\
S_F^{ij}( {\bf r}_1, {\bf r}_2; -\omega )
& = & 
2 \hbar \bar{n}_{\rm th}
\mathop{\rm Im}\, G_{ij}( {\bf r}_1, {\bf r}_2; \omega ).
\end{eqnarray}
At zero temperature, $\bar{n}_{\rm th} = 0$, and only the first line survives.
The relaxation dynamics is then entirely due to spontaneous decay, induced
by the vacuum fluctuations of the force field. Heating processes are
suppressed. At high temperature, 
$\bar{n}_{\rm th} \gg 1$, the fluctuation spectrum becomes independent
of the sign of $\omega$. In the master equation, decay and excitation 
rates are then nearly the same.


\section{Asymptotic expansion of electromagnetic field spectra} 
\label{a:quasi-static-limit}

\subsection{Electric field}
\label{a:E-quasi-static-limit}

We outline here the asymptotic expansion for the coefficients
$g_{\Vert,\perp}( k z )$ that characterize the electric field
fluctuations~(\ref{eq:prox-E-field}) in the near field $k z \ll 1$ 
of the surface.

The inspection of the integrals~(\ref{eq:g-tensor}) shows that
the exponential $e^{ 2 i k z v } \approx e^{ - 2 k z u }$
decreases on a large scale $u \sim 1/(k z ) \gg 1$. 
On the other hand, the
other factors in the integrands increase as powers of $u$.
The value of the integral is thus dominated by values $u \sim u_{\max}$
around the maximum $u_{\max} \sim 1/( k z ) \gg 1$. It is therefore
accurate to use asymptotic expansions of the Fresnel coefficients
for large $u \gg 1$. The asymptotic form of the coefficients
depends, however, on whether $u^2_{\max}$ is smaller or larger
than the magnitude $|\varepsilon|$ of the dielectric constant.
These two regimes are discussed in the following. Their physical
significance follows from the relation~(\ref{eq:eps-and-delta})
between $\varepsilon$ and the skin depth $\delta$.
 
The limit $1  \ll |\varepsilon|^{1/2} \ll u$
corresponds to a distance small compared to the skin depth,
$ z \ll \delta \ll \lambdabar$. In this regime, 
we get the following asymptotic expressions for the Fresnel coefficients~%
(\ref{eq:Fresnel-rsp})
\begin{eqnarray}
&& r_p(u) \to \frac{\varepsilon - 1}{\varepsilon + 1}, 
\nonumber
\\
&& r_s(u) \to \frac{\varepsilon - 1}{4 u^2}. 
\end{eqnarray}
The integrals~(\ref{eq:g-tensor}) for the tensor elements are then 
evaluated to 
\begin{eqnarray}
&& g_\Vert( k z ) \approx
\frac{ 3 }{ 16 (k z)^3 }
\mathop{\rm Im}\, \frac{ \varepsilon - 1 }{ \varepsilon + 1 }
\approx
\frac{ 3 \delta^2 }{ 16 k z^3 }
,
\nonumber
\\
&& g_\perp( k z ) \approx
2 g_\Vert( k z ).
\label{eq:g-short-asymptotics}
\end{eqnarray}
In the opposite limit of a small skin depth, 
i.e.\ $ \delta \ll z \ll \lambdabar $, we have
$1 \ll u \ll |\varepsilon|^{1/2}$,
and the reflection coefficients show the asymptotic behavior
\begin{eqnarray}
&& r_p(u) \to 1 + \frac{2i}{u\sqrt{\varepsilon}}, 
\nonumber
\\
&& r_s(u) \to -1 + \frac{2iu}{\sqrt{\varepsilon}}.  
\label{eqa:rsp-asymptotics}
\end{eqnarray}
This yields tensor elements of the form
\begin{eqnarray}
&& g_\Vert( k z ) \approx
\frac{ 3 }{ 4 (k z)^2 }
\mathop{\rm Re}\, \frac{ 1 }{ \sqrt{ \varepsilon } }
\approx
\frac{ 3 \delta }{ 8 k z^2 }
,
\nonumber
\\
&& g_\perp( k z ) \approx
g_\Vert( k z ).
\label{eq:g-intermediate-asymptotics}
\end{eqnarray}
The regimes~(\ref{eq:g-short-asymptotics},%
\ref{eq:g-intermediate-asymptotics}) are readily combined into the
interpolation formula~(\ref{eq:result-g-tensor}).

In the limit of a perfectly conducting ({\sc pc}) surface 
($\varepsilon \to \infty$), the skin depth $\delta$ vanishes, and 
the reflection coefficients~(\ref{eq:Fresnel-rsp}) 
are equal to $r_{p,s} = \pm 1$ (cf.\ eq.(\ref{eqa:rsp-asymptotics})). 
The integrals~(\ref{eq:g-tensor}) 
may be evaluated explicitly, and one gets
\begin{eqnarray}
\mbox{\sc pc}:\quad
g_\Vert( k z ) & = & \frac32
\left(
\frac{ \sin 2 k z }{ (2 k z)^3 } -
\frac{ \cos 2 k z }{ (2 k z)^2 } -
\frac{ \sin 2 k z }{ 2 k z }
\right)
,
\nonumber\\
g_\perp( k z ) & = & 3
\left(
\frac{ \sin 2 k z }{ (2 k z)^3 } -
\frac{ \cos 2 k z }{ (2 k z)^2 } 
\right)
.
\label{eq:pc-asymptotics}
\end{eqnarray}
Note that these functions have finite limiting values at $z \to 0$,
which is different from the behavior~(\ref{eq:g-short-asymptotics}) 
above a surface with a finite conductivity.

\subsection{Magnetic field}
\label{a:B-quasi-static-limit}

The asymptotic evaluation of the coefficients $h_{\Vert,\perp}( k z )$
for the magnetic field spectrum~(\ref{eq:prox-B-field}) proceeds
similar to the case of the electric field.
 
For a skin depth larger than the trap distance, we expand the
reflection coefficients in the regime 
$1 \ll |\varepsilon|^{1/2} \ll u$. 
The asymptotics of the tensor elements~(\ref{eq:def-h-elements}) is then
given by
\begin{eqnarray}
&& z \ll \delta \ll \lambdabar: 
\nonumber
\\
&& h_\Vert( k z ) \approx 
\frac{ 3 }{ 32 k z }
\mathop{\rm Im}\, \frac{ (\varepsilon - 1) (\varepsilon + 5) }{
\varepsilon + 1 }
\approx
\frac{ 3 }{ 16 k^3 \delta^2 z }
,
\nonumber
\\
&& h_\perp( k z ) \approx 
\frac{ 3 }{ 16 k z }
\mathop{\rm Im}\, (\varepsilon - 1)
\approx
2 h_\Vert( k z )
.
\label{eq:h-short-asymptotics}
\end{eqnarray}
We used the approximation $|\varepsilon| \gg 1$
appropriate for a good conductor.

In the opposite limit of a small skin depth, we find
\begin{eqnarray}
&& \delta \ll z \ll \lambdabar:
\nonumber
\\
&& h_\Vert( k z ) \approx 
\frac{ 9 }{ 16 (k z)^4 }
\mathop{\rm Re}\, \frac{ 1 }{ \sqrt{ \varepsilon } }
=
\frac{ 9 \delta }{ 32 k^3 z^4 }
,
\nonumber
\\
&& h_\perp( k z ) \approx 
2 h_\Vert( k z )
.
\label{eq:h-intermediate-asymptotics}
\end{eqnarray}
Both expressions~(\ref{eq:h-short-asymptotics},%
\ref{eq:h-intermediate-asymptotics}) are reproduced by the 
interpolation formula~(\ref{eq:result-h-tensor}).


\end{document}